# What Is a System? An Interaction-Based Account of Structure–Behavior Coalescence in General Systems Theory

William S. Chao

**Abstract** — The question of what constitutes a system remains fundamental to General Systems Theory. Existing definitions commonly characterize a system in terms of elements, relationships, boundaries, functions, or interactions, but these perspectives do not always provide a unified account of how system structure and system behavior constitute one another. This paper proposes Structure–Behavior Coalescence (SBC) as an interaction-based account of what a system is. From the SBC perspective, a system is not merely a collection of elements or relationships, nor is it adequately characterized by behavior considered independently of structure. Rather, a system is a structured entity whose behavior arises through interactions among its constituent entities, with those interactions simultaneously contributing to both its structural organization and behavioral realization. The paper develops this perspective by distinguishing system structure, interaction, and behavior while treating them as inseparable aspects of a unified system representation. It argues that interactions provide the essential link through which structural relationships become behavioral processes and through which behavioral processes reveal and instantiate system structure. This perspective provides a basis for representing systems in a manner that maintains consistency between what a system is and what a system does. The paper further considers the implications of SBC for General Systems Theory, particularly for system identity, system boundaries, behavioral emergence, and the representation of complex systems. SBC is presented as a general conceptual foundation that complements existing systems theories by placing the coalescence of structure and behavior through interaction at the center of the definition of a system.

## 1. Introduction

The question "what is a system?" lies at the foundation of systems science. Since the development of General Systems Theory, systems have been understood as organized wholes whose properties arise from interactions among components rather than from the components themselves in isolation [1–3]. This perspective has provided a powerful conceptual shift away from reductionist thinking and has influenced a wide range of disciplines, including engineering, biology, organizational science, and computer science.

Despite this shared foundation, most contemporary systems representations continue to rely on a persistent conceptual separation between structure and behavior. Structural descriptions typically focus on system components and their relationships, while behavioral descriptions focus on state changes, processes, or event sequences over time. This distinction is deeply embedded in modeling practices, including multi-view frameworks such as UML [4–5] and SysML [6], as well as many formal and semi-formal methods used in systems engineering [7–8].

While this separation is useful for managing complexity, it introduces a recurring difficulty: structural and behavioral views are not intrinsically aligned. Instead, their consistency must be maintained through additional mechanisms such as traceability links, transformation rules, or external consistency constraints [9–10]. As systems become more complex and increasingly socio-technical in nature, these alignment requirements become a significant source of modeling overhead and conceptual fragmentation [11].

This paper argues that this difficulty is not merely a technical limitation but reflects a deeper conceptual assumption: that structure and behavior are fundamentally distinct aspects of a system. Under this assumption, systems must be modeled through multiple independent perspectives that are later reconciled. However, this raises a more fundamental question: whether the separation itself is necessary for system description, or whether it is primarily a consequence of representational tradition.

To address this question, this paper introduces an interaction-based account of systems grounded in the principle of Structure–Behavior Coalescence (SBC). The central idea is that systems are not primarily composed of static structures that subsequently exhibit behavior. Instead, systems are constituted through interactions, and both structure and behavior can be understood as arising from the same interaction-based foundation. Interactions are treated as the atomic units of system specification, simultaneously encoding participating entities and executed operations.

From this perspective, structure is interpreted as the stabilization of interaction patterns over time, while behavior is understood as the unfolding of these interaction-based compositions. Structure and behavior are therefore not independent modeling dimensions, but co-emergent interpretations of a single underlying interaction-based reality.

To make this perspective operational, the paper introduces a minimal interaction-based grammar for system specification, in which system processes are constructed through sequential, alternative, and concurrent composition of interaction primitives. This formal core is intentionally lightweight, serving as a representational substrate rather than a full process algebra, and is used to illustrate how both structural and behavioral views can be derived from a unified description.

The contribution of this paper is therefore twofold. First, it proposes a conceptual shift in how systems are defined, moving from component-based and dual-view representations toward interaction-based system constitution. Second, it demonstrates how this shift provides a unified basis for understanding structure and behavior as coalesced aspects of system organization.

The remainder of the paper is organized as follows. Section 2 reviews relevant background in General Systems Theory and highlights limitations of structure–behavior separation. Section 3 introduces the interaction-based foundation of system representation and presents the minimal SBC grammar. Section 4 formulates the Structure–Behavior Coalescence principle. Section 5 discusses system identity, emergence, and representation under the proposed framework. Section 6 introduces a view derivation mechanism that connects interaction-based representations to structural and behavioral projections. Section 7 situates the approach within classical systems theory. Section 8 presents a worked system example using the proposed formalism. Finally, Section 9 concludes the paper.

## 2. Background and Related Work

### 2.1 General Systems Theory and the Concept of Organization

General Systems Theory (GST) introduced the foundational idea that systems should be understood as organized wholes rather than as collections of isolated parts [1–3]. In this tradition, pioneered by Bertalanffy, the emphasis is placed on the interactions and interdependencies among components as the source of system-level properties [2–3, 14–15]. This marked a significant departure from reductionist approaches, shifting attention toward organization, wholeness, and system-level phenomena.

However, despite this conceptual shift, GST does not fully formalize how organization itself should be represented. Systems are still typically described in terms of components and relations, where interactions are treated as secondary structures defined over pre-existing entities. As a result, the notion of organization remains largely descriptive rather than generative.

This paper extends the GST intuition that organization is fundamental, but seeks to make interaction itself the primary generative construct of system representation.

### 2.2 Cybernetics and Interaction-Based Control

Cybernetics, particularly in the work of Ashby and Beer, provides a complementary perspective by focusing on regulation, control, and communication in complex systems. Systems are modeled as feedback structures in which behavior emerges from interaction loops between system components and their environment [12–15].

While cybernetics emphasizes dynamic interaction, it still assumes a separation between system structure (components and channels of communication) and system behavior (state evolution and control outcomes) [12, 16]. The modeling of feedback, regulation, and adaptation is therefore built on an underlying representational duality between structure and dynamics.

SBC departs from this duality by treating interaction as the only primitive construct, from which both structural and behavioral interpretations are derived.

### 2.3 Process-Oriented and Dynamic Systems Perspectives

Process-oriented approaches in philosophy and systems theory further challenge static conceptions of systems by treating processes as more fundamental than objects [17]. In these perspectives, systems are seen as ongoing processes of becoming rather than fixed assemblies of components [18–19].

Such approaches align closely with the SBC viewpoint in their emphasis on dynamism and relationality [17–19]. However, many process-based frameworks remain conceptually informal or lack a minimal compositional structure for representing system construction.

SBC addresses this gap by introducing a minimal interaction-based grammar that supports compositional system construction without reintroducing a strict structure–behavior separation.

### 2.4 Multi-View Modeling and Structure–Behavior Separation

In systems engineering and software engineering practice, multi-view modeling frameworks such as UML and SysML explicitly separate structural and behavioral views [4–6]. Structural diagrams describe components and their relationships, while behavioral diagrams capture interactions, state transitions, and workflows.

Although this separation improves modularity and manageability, it introduces a well-known consistency problem: structural and behavioral models must be kept aligned through external mechanisms such as traceability links, transformation rules, or consistency constraints [9–11].

This reliance on external alignment suggests that the structure–behavior distinction may be a property of representation rather than a property of systems themselves.

### 2.5 Limitations of Existing Approaches

Across General Systems Theory, cybernetics, process-based modeling, and multi-view engineering, a common pattern emerges: systems are described through multiple representational layers that separate structure from behavior [9–11]. While this separation is useful for abstraction and design, it introduces several limitations.

Structural and behavioral models often evolve independently, increasing the risk of inconsistency [8–11]. Traceability must be explicitly maintained, introducing additional modeling overhead. Consistency checking is typically performed post hoc rather than being inherent to the modeling formalism. Most importantly, there is no unified primitive from which both structure and behavior can be directly derived.

These limitations suggest that the challenge is not only technical but also conceptual: existing approaches assume a fundamental separation between structure and behavior that may not be necessary at the level of system constitution.

### 2.6 Motivation for an Interaction-Based Reformulation

The limitations identified above motivate a shift toward an interaction-based foundation of systems. Instead of treating interactions as relationships between pre-defined entities or as transitions over predefined states, interactions can be taken as the primary building blocks of system representation.

This paper develops this idea through the principle of Structure–Behavior Coalescence, in which system structure and system behavior are not treated as separate constructs but as co-emergent interpretations of interaction-based composition. This perspective provides the basis for the formal and conceptual framework introduced in the following sections.

## 3. Interaction-Based Foundation of System Representation

This section introduces the minimal formal basis of Structure–Behavior Coalescence. The objective is to define a compact interaction-based grammar that supports the construction of system processes and provides a representational basis from which structural and behavioral interpretations can later be derived.

### 3.1 Interaction as the Primitive of System Representation

In SBC, the minimal syntactic unit of system specification is the interaction. An interaction $a_1$ denotes an atomic element of system activity [20–21].

Each interaction is treated as a syntactic symbol without internal semantic or referential interpretation. No interpretation function is defined at the level of Section 3. Any interpretation over interactions is introduced only at the level of derived projections in Section 4. These elements are not treated as separate modeling constructs, but as aspects associated with the interpretation of an interaction within a system specification.

Let $A = \{a_1, a_2, a_3, \ldots\}$ denote the set of atomic interactions in the system domain.

### 3.2 Minimal System Grammar

A system process P is defined by the following grammar: $P ::= 0 \mid (g, a_1) \bullet P \mid P + P \mid P \parallel P \mid ITG$, where 0 denotes termination or an inactive system state, g is a Boolean guard condition controlling execution, and $a_1 \in A$ is an atomic interaction [20–22]. The expression $(g, a_1) \bullet P$ denotes guarded prefix sequencing, meaning that if the condition g holds, interaction $a_1$ occurs and the system continues as process P. The expression P + P denotes alternative composition, representing behavioral branching, while $P \parallel P$ denotes concurrent composition of parallel interaction-based processes. ITG denotes an Interaction Transition Graph representation that is *structurally aligned with* the process grammar and provides an equivalent representation up to graph–process translation semantics defined in [20–22].

Operator precedence is defined such that $\bullet$ has higher precedence than +, and + has higher precedence than $\parallel$. This grammar defines a compositional system in which valid processes are constructed exclusively from interaction primitives and composition operators.

### 3.3 Guarded Prefix Sequencing of Interactions

Guarded prefix sequencing $(g, a) \bullet P$ specifies that interaction a is enabled when condition g holds, after which the system continues as process P. For example, $(g, a_1) \bullet (g', a_2) \bullet 0$ represents a process in which $a_1$ precedes $a_2$ subject to their respective guard conditions.

In this framework, sequencing is interpreted as a compositional ordering constraint over interactions within a process definition.

### 3.4 Choice and Behavioral Branching

Alternative composition $P_1 + P_2$ specifies that a process may evolve according to one of multiple possible interaction paths. For example, $(g_1, a_1) \bullet P_1 + (g_2, a_2) \bullet P_2$ indicates that execution proceeds along the branch whose guard condition is satisfied.

Choice is represented as a structural composition operator at the level of process syntax, rather than as an external decision mechanism.

### 3.5 Parallel Composition

Parallel composition $P_1 \parallel P_2$ specifies concurrent execution of two processes within the same system specification.

The resulting compositions admit multiple valid interaction orderings induced by syntactic structure of interactions from the composed processes, while preserving their structural origin within the composition.

This construction follows standard notions of concurrency in process-algebraic systems [20–21, 23].

### 3.6 Interaction Transition Graph

The Interaction Transition Graph (ITG) provides a graphical encoding of syntactic process structure. In this representation, nodes are syntactic graph elements derived from interaction prefixes, while directed edges are labeled with interactions $a \in A$. Transitions between nodes are labeled by interactions and annotated with guard expressions associated with the corresponding interactional steps.

The ITG denotes an Interaction Transition Graph representation that is consistent with the process grammar and has a syntactic correspondence with process grammar via structural encoding [20–23].

It serves as an alternative representation of the same syntactic structure defined by the process grammar. It does not extend the expressive class of the process grammar and is interpreted via a standard graph–process correspondence.

### 3.7 Interpretive Role of the Interaction Model

The grammar in Section 3 defines only the formation rules for valid system processes and does not assign any interpretation to such processes at the syntactic level. All semantic interpretation is deferred to Section 4, where structural and behavioral interpretations are introduced. Consequently, multiple analytical readings are deferred to Section 4. The specific interpretation adopted in SBC is defined in Section 4, where projection functions Struct(P) and Beh(P) are defined as projections over the interaction-based process specification. All constructs in Sections 3–5 are defined purely at the syntactic level of process formation; no observational, semantic, or state-based interpretation is assumed in the definition of P. All structural and behavioral properties are introduced only through the projection functions defined in Section 4.

## 4. Structure–Behavior Coalescence Principle

This section introduces the Structure–Behavior Coalescence principle as an interpretive principle over the interaction-based system representation defined in Section 3. While Section 3 defines a minimal syntactic framework for constructing system processes, this section specifies how structural and behavioral interpretations are obtained from that framework.

### 4.1 From Interactional Syntax to System Interpretation

The interaction-based grammar introduced in Section 3 defines a space of valid system processes P constructed from atomic interactions and compositional operators. Over this process space, SBC introduces structural and behavioral interpretations derived from a common interaction-based specification. These interpretations define the projection basis used in Section 4. Such interaction-centered process descriptions are consistent with established interaction-based and process-oriented representations developed in concurrency theory and systems modeling [20–22]. However, this syntactic level does not distinguish between structural and behavioral descriptions.

The Structure–Behavior Coalescence principle is introduced at this level as a formally defined projection function over interaction-based processes. It specifies how different analytical views can be systematically obtained from a single interaction-based representation without introducing additional modeling primitives.

### 4.2 Structural Interpretation over Interaction Structure in P

Within SBC, structural interpretations are obtained by analyzing syntactic relations over interaction structure in P.

Rather than being explicitly specified as a separate structural model, structure is derived from syntactic relations induced by interaction occurrences induced by interaction occurrences in the process execution space. In particular, structural relationships correspond to stable or repeated associations between entities that arise through their involvement in interaction structures in P. This perspective is consistent with systems-theoretic accounts that associate system structure with patterns of organization and relational interdependence rather than with isolated components [1–3, 14-15].

This interpretation does not require structure to be defined independently of the process. Instead, structure is treated as a derived view obtained from the organization of interactions in P.

### 4.3 Behavioral Interpretation over Interaction Structure in P

Behavioral interpretations are obtained by analyzing the ordered occurrence of interactions within the same process P.

In this view, behavior corresponds to the ordered sequences of interactions induced by the compositional operators defined in the grammar, including sequencing, choice, and parallel composition. Syntactic derivation traces of P yield different interaction sequences, which together form the behavioral interpretation of the system. This interpretation is consistent with process-algebraic and interaction-based models in which system evolution is represented through structured compositions of observable interaction events [20–22].

Behavior is therefore not introduced as a separate modeling layer, but as a derived view of interaction structure in P focused on execution order and process evolution.

### 4.4 The Structure–Behavior Coalescence Principle

The Structure–Behavior Coalescence principle states that structural and behavioral interpretations arise from the same syntactic interaction-based process P defined in Section 3.

Given a process P, both structure and behavior are obtained through different analyses of the same interaction-based process: structure through Struct(P) defined as a mapping over syntactic relations induced by interaction prefixes in P, and behavior through Beh(P) defined by ordering relations over syntactic derivation traces of P. These interpretations do not require independent specification or external alignment at the level of representation, since they are derived from a shared syntactic origin. This differs from conventional multi-view modeling approaches, where structural and behavioral descriptions are typically maintained as separate artifacts and subsequently reconciled through traceability, transformation, or consistency mechanisms [4–11].

Accordingly, SBC does not introduce structure and behavior as primitive modeling constructs. Instead, it provides a single interaction-based representation from which both views can be systematically derived as complementary interpretations.

For convenience, the structural interpretation of a process P is denoted Struct(P), and the behavioral interpretation is denoted Beh(P). These notations refer to derived views obtained from the same interaction-based process specification and do not constitute additional primitives of the formalism.

### 4.5 Conceptual Consequence

The primary implication of this formulation is representational rather than ontological. Once system specification is expressed as an interaction-based process, structural and behavioral descriptions no longer need to be introduced as independent modeling primitives.

Instead, they function as two analytically distinct but formally grounded interpretations of the same underlying process P. This reframes the structure–behavior relationship as a property of projection consistency over a shared syntactic process P, rather than as a relation between separately defined models. In this respect, SBC remains compatible with the systems-theoretic emphasis on organization, interdependence, and system-level coherence, while proposing a different representational basis for expressing those concepts [1–3].

## 5. System Identity, Emergence, and Representation

All references to structural and behavioral forms in this section refer to the projection functions introduced in Section 4 and are not additional primitives of the formal system.

Building on the interpretive framework introduced in Section 4.5, this section examines system identity, emergence, and representation within the interaction-based modeling framework. These notions are not introduced as independent primitives, but are instead interpreted as analytical consequences of applying the structural and behavioral projections defined in Section 4 to a unified interaction-based specification defined in Section 3.

Within this setting, system identity, emergent properties, and representation are not treated as separate conceptual layers of the formalism. Instead, they are considered distinct analytical perspectives arising from a single interaction-based process representation, consistent with the multi-view motivation in systems engineering literature [4–6, 9–11].

### 5.1 System Identity in the Interaction-Based Framework

In classical systems theory, system identity is often associated with persistence of components, structural organization, or behavioral equivalence under transformations [1–3, 7–8]. These formulations typically assume a representational separation between structural and behavioral descriptions, with identity defined relative to one or both.

Within the interaction-based framework, system identity is interpreted as a property of interaction structure in P as defined in Section 3 and analyzed through the interpretive mechanisms of Section 4. Specifically, identity is analyzed with respect to the structural and behavioral interpretations introduced in Section 4 over a process P and its induced interaction structure in P, rather than to a fixed set of components or states.

From this perspective, system comparison may be analyzed in terms of correspondence between interactional process structures as interpreted through the structural and behavioral views introduced in Section 4, including sequencing, branching structure, and concurrency structure, as induced by the process grammar. This notion is consistent with relational and organization-centered interpretations of systems found in General Systems Theory, where system identity is associated with organized relations rather than isolated elements [1–3].

However, in contrast to traditional formulations, these relations are not assumed to exist independently of the system specification. Instead, they are derived from the interaction-based representation itself.

Accordingly, system identity in SBC is interpreted as arising from interaction structure in P rather than being defined over separately specified structural or behavioral models.

### 5.2 Emergence from Interaction Structure in P

Emergence is commonly defined as the appearance of system-level properties that are not directly attributable to individual components but arise from their interactions and organization [1–3, 14]. In many classical accounts, emergence is treated as an explanatory layer in addition to structural or behavioral descriptions.

Within SBC, emergence is interpreted at the level of interaction-based representation and its derived interpretations (Section 4). Since system processes are defined through compositions of interactions, emergence refers to properties that are not explicit in the syntactic process P but become observable through the structural and behavioral projections defined in Section 4.

Local interactions contribute to execution-level behavior, while recurring configurations of interaction structure in P give rise to higher-level regularities observable in derived views. These regularities correspond to what is typically described as emergent phenomena in systems theory.

This interpretation remains compatible with cybernetic and systems-theoretic accounts of emergence [12–16], where emergence is associated with feedback, self-organization, and global behavioral regularities. However, SBC does not introduce emergence as an additional modeling layer beyond interaction-based composition and its interpretations.

### 5.3 Structural and Behavioral Representations as Derived Views

In multi-view systems engineering, structural and behavioral models are typically maintained as separate artifacts that require alignment through traceability, transformation, or consistency mechanisms [4–6, 9–11]. This reflects a modeling practice in which structure and behavior are treated as independent representational views.

Within SBC, structural and behavioral representations are not introduced as independent models. Instead, they are obtained by applying the structural and behavioral interpretations introduced in Section 4 to the interaction-based process P.

The structural interpretation emphasizes patterns of interactional participation across entities, corresponding to stable relational organization induced by repeated interaction occurrences. This is consistent with systems-theoretic perspectives in which structure is understood as persistent relational organization rather than component aggregation [1–3, 14].

The behavioral interpretation emphasizes the ordered occurrence of interactions induced by the process structure, including sequencing, choice, and concurrency. This aligns with process-algebraic and concurrency-theoretic models in which system evolution is represented as structured compositions of observable interaction events [19–21].

Since both interpretations are derived from the same interaction-based specification, no additional reconciliation is required at the representational level. However, this does not eliminate the usefulness of multiple views in practice; rather, it reframes them as projections of a shared formal source, consistent with established multi-view modeling motivations [4–6, 9–11].

### 5.4 Consistency as a Derived Property

In multi-view systems engineering, consistency between structural and behavioral models is typically enforced through external mechanisms such as traceability links, transformation rules, or constraint systems [9–11]. This necessity arises because the models are defined independently.

In SBC, consistency is not introduced as an additional modeling constraint. Instead, it is treated as a property of alignment between the structural and behavioral projections of the same interaction-based process P, as specified in Sections 3 and 4.

From this perspective, consistency is not an externally imposed relation between models but a property of the derivation process itself. Apparent inconsistencies arise when structural or behavioral interpretations are constructed independently of the underlying interaction-based specification.

This interpretation aligns with prior observations in multi-view modeling research regarding the role of traceability and model alignment [4–6, 9–11], but reframes these mechanisms as necessary primarily under representational separation.

### 5.5 Implications for System Representation

The interaction-based interpretation of system identity, emergence, and representation leads to a unified perspective on system modeling.

First, system identity is associated with properties of the structural and behavioral interpretations of a process rather than component-level or state-level equivalence in isolation, consistent with relational views in General Systems Theory [1–3].

Second, emergence is interpreted as a property of interaction structure in P observable through structural and behavioral projections, consistent with classical systems and cybernetic accounts of self-organization and global behavior [12–15].

Third, structural and behavioral representations are treated as derived projections rather than independent modeling artifacts, consistent with multi-view modeling motivations in systems engineering [4–6, 9–11].

Overall, SBC does not replace existing modeling approaches. Instead, it provides a unified representational basis from which multiple system views can be systematically derived without introducing additional primitive modeling constructs.

## 6. View Derivation Framework

This section specifies how structural and behavioral system views are obtained from a single interaction-based representation. Within the Structure–Behavior Coalescence framework, system descriptions are not introduced as independent models but are derived from a unified interaction-based specification P defined by the grammar in Section 3.

The purpose of this framework is to make explicit the derivation structure underlying different analytical views, without introducing additional modeling primitives beyond interaction-based composition. Accordingly, structural and behavioral representations are treated as projections rather than separately specified artifacts.

### 6.1 From Unified Representation to Multiple Views

Let P denote a system process constructed from atomic interactions using sequential, alternative, and concurrent composition. System views are obtained through projection mappings over P, rather than through independent model construction. In this setting, a view corresponds to a mapping from interaction-based structure to a specific analytical perspective.

This formulation preserves the role of multiple views in systems engineering while ensuring that they share a common representational origin [4–6, 9–11].

### 6.2 Structural View Derivation

The structural view is obtained by examining patterns of interaction-based participation across entities within the system process P. In particular, structure is defined by Struct(P), which captures syntactic relations induced by P.

From this perspective, structural organization is not introduced as an independent model but is defined as a projection over the syntactic specification of P across executions of P. This interpretation is consistent with systems-theoretic accounts in which structure is associated with stable relational organization rather than isolated components [1–3, 14].

### 6.3 Behavioral View Derivation

The behavioral view is derived from the compositional ordering of interactions within the system process P. This includes sequencing induced by prefix composition, branching induced by alternative composition, and concurrency induced by parallel composition.

Accordingly, behavior is represented as the ordering structure induced by syntactic derivation traces of P governed by the process structure of P. This interpretation aligns with established process-based and interaction-based models in which system behavior is represented as ordering relations over interaction sequences derived from P rather than state-centric transitions [19–21].

### 6.4 Trace-Based View

In addition to structural and behavioral views, a trace-based interpretation may be obtained by considering particular execution paths of the system process P. A trace corresponds to a linear or partially ordered sequence of interaction occurrences obtained from a selected syntactic derivation trace of P.

This provides a derived syntactic perspective over interaction sequences in P that is particularly relevant for analysis and verification, while remaining fully grounded in the same interaction-based specification.

### 6.5 Unification through Shared Derivation Source

A key property of this framework is that all views are derived from the same underlying specification P. As a result, structural and behavioral descriptions do not require external synchronization or alignment at the representational level.

However, this should be understood as a property of shared derivation rather than a guarantee of agreement between independently constructed models. When views are constructed independently of the syntactic structure of P, additional reconciliation mechanisms may still be required at the level of modeling practice.

Accordingly, the role of the framework is not to eliminate multiple views, but to ensure that when such views are derived as projections from the same syntactic specification of P, their relationship is determined by that specification.

## 7. Positioning within Classical Systems Theory

This section situates Structure–Behavior Coalescence within established traditions in systems science, including General Systems Theory, cybernetics, process-oriented philosophy, and multi-view systems engineering. The purpose is not to replace these frameworks, but to clarify how SBC relates to and reorganizes existing system-theoretic assumptions through an interaction-based representation.

7.1 Continuity with General Systems Theory

General Systems Theory (GST) emphasizes that system properties arise from organized wholes rather than isolated components [1–3]. This perspective already places interaction and interdependence at the center of system explanation, even if it does not fully formalize a minimal representational primitive for constructing systems.

SBC remains consistent with this foundational intuition. It retains the view that system properties are produced by organized relations among elements, while making interaction the explicit unit through which such organization is represented.

7.2 Relation to Cybernetic Approaches

Cybernetic approaches emphasize feedback, regulation, and communication as central mechanisms in system behavior [12–15]. These models already highlight the importance of interaction-based coupling between system components and their environment.

However, cybernetic formalisms typically maintain a separation between structural organization and behavioral dynamics. SBC differs in that both aspects are represented through a single interaction-based specification, rather than through separate structural and control layers.

7.3 Relation to Process-Oriented System Perspectives

Process-oriented approaches in systems theory and philosophy treat processes as more fundamental than static entities [16–18]. These approaches align with SBC in emphasizing dynamic relationality and ongoing transformation.

SBC complements these perspectives by introducing a minimal compositional grammar for interaction-based construction, allowing system processes to be specified in a structured and analyzable form.

7.4 Relation to Multi-View Systems Engineering

In systems engineering practice, structural and behavioral views are typically separated to manage complexity, as reflected in UML and SysML-based modeling approaches [4–6]. This separation introduces well-known challenges in consistency management and traceability [9–11].

SBC addresses this issue by treating structural and behavioral descriptions as derived projections of a single interaction-based specification. This does not remove the usefulness of multi-view modeling, but reframes it as a consequence of interpretive projection rather than independent model construction.

7.5 SBC as a Minimal Representational Reorganization

SBC should be understood as a minimal reorganization of system representation rather than as a replacement of existing systems theories. Its central contribution is to make interaction explicit as the shared representational basis from which structural and behavioral interpretations can be derived.

This reorganization preserves compatibility with established system-theoretic concepts while providing a unified basis for derivation and analysis.

7.6 Why Interaction?

The central design choice in Structure–Behavior Coalescence is the selection of interaction as the primitive unit of system representation. This choice is not intended as an axiomatic assumption but as a consequence of comparative representational sufficiency across standard system-theoretic primitives.

If components are taken as primitive, systems are represented as collections of entities, but such representations do not directly encode how system organization emerges without introducing additional relational or behavioral constructs. If behavior is taken as primitive, system descriptions rely on trajectories or event sequences, but these presuppose a domain of entities and a notion of state. If relations are taken as primitive, systems are described through static associations, which do not directly capture execution or temporal evolution. If processes are taken as primitive, system descriptions often implicitly rely on either underlying entities or predefined interaction rules to explain what is being processed and how transitions are structured.

Interaction differs in that it functions as the minimal syntactic unit from which both structural and behavioral interpretations can be derived through the projection framework introduced in Section 4. In this sense, interaction serves as a minimal syntactic unit of representation from which structural and behavioral interpretations can be derived via Section 4.

From this perspective, interaction is not privileged as a metaphysical foundation, but selected as the minimal syntactic unit sufficient to support derivation of structural and behavioral projections within SBC. Accordingly, SBC does not eliminate alternative system-theoretic notions; instead, it re-expresses them as derived interpretations over the syntactic specification of P, where components are represented through their involvement in interactions defined within P, relations correspond to stable patterns induced by interaction occurrences in P, and behavior corresponds to ordering relations over syntactic derivation sequences.

## 8. Worked System Example

This section presents a minimal worked example to illustrate how a complete system can be constructed, interpreted, and analyzed using the Structure–Behavior Coalescence (SBC) framework. The goal is not to model a full-scale application system, but to demonstrate how interaction-based composition generates both structural and behavioral interpretations from a single representation.

8.1 System Description: Smart IoT Monitoring Scenario

Consider a simplified smart monitoring system consisting of three roles: a Sensor node (S), a Processing unit P, and an Alert service (A). The system detects an event, processes it, and triggers an alert if necessary.

We define the following atomic interactions: $a_1$ represents Sensor S sending data to Processor P, $a_2$ represents Processor P

evaluating the data, $a_3$ represents Processor P sending an alert request to Alert service A, and $a_4$ represents Alert service A issuing a notification. Let $A = \{a_1, a_2, a_3, a_4\}$.

8.2 Interaction-Based System Specification

The system is specified using the SBC grammar introduced in Section 3. Let P be a system process defined as: $P ::= (\text{true}, a_1) \bullet (\text{true}, a_2) \bullet ((g, a_3) \bullet (\text{true}, a_4) \bullet 0 + (\neg g, a_2) \bullet 0)$. In this specification, $a_1$ and $a_2$ represent mandatory sensing and evaluation steps, g is a condition representing detection of an abnormal event, $a_3$ and $a_4$ represent alert generation, and $\neg g$ represents the normal (non-alert) operational path.

This specification defines two possible interaction-based evolutions: a normal processing path without alert generation and an alert generation path.

8.3 Behavioral Interpretation

From a behavioral perspective, the system generates two possible interaction traces [19–20, 22]. The normal trace is given by $a_1 \rightarrow a_2$, while the alert trace is given by $a_1 \rightarrow a_2 \rightarrow a_3 \rightarrow a_4$. These traces are not separately defined behaviors but are directly derived from the interaction-based composition of P.

Behavior is therefore not an additional model but a behavioral interpretation derived from the system process.

8.4 Structural Interpretation

From a structural perspective, system organization is derived from patterns associated with interaction occurrences [1–3]. The Sensor S participates in $a_1$, the Processor P participates in $a_1$, $a_2$, and $a_3$, and the Alert service A participates in $a_4$.

From these recurring interaction patterns, structural relationships are derived, including $S \rightarrow P$ via $a_1$, $P \rightarrow P$ via self-processing in $a_2$, and $P \rightarrow A$ via $a_3$. Thus, system structure is not explicitly defined but is derived from the distribution of interactions across entities.

8.5 Structure–Behavior Coherence

Both structural and behavioral interpretations originate from the same process P. Behavior is obtained by unfolding interaction sequences, while structure is obtained by analyzing recurring interaction patterns within P.

No additional mapping or traceability mechanism is required to relate the two views, unlike conventional multi-view modeling approaches that typically require explicit alignment, traceability, or transformation mechanisms [4–11]. Their consistency follows from the fact that both are derived from a single interaction-based specification. This illustrates the core SBC claim that structure and behavior are complementary interpretations of interaction-based composition rather than independent modeling layers.

## 9. Conclusion

This paper has proposed an interaction-based account of systems grounded in the principle of Structure–Behavior Coalescence. The central idea is that a system is not fundamentally a collection of components, nor a dual description of structure and behavior, but an organized composition of interactions from which both structural and behavioral interpretations are derived.

By treating interactions as the atomic unit of system representation, SBC removes the need for a prior separation between structural and behavioral modeling. Structure is reinterpreted as stabilized patterns derived from interaction occurrences, while behavior is understood as the temporal unfolding of interaction-based composition. These two aspects are not independently specified but are derived from the same underlying process.

To support this view, a minimal interaction-based grammar was introduced, providing a compact but expressive mechanism for constructing system processes through sequential, alternative, and concurrent composition. The worked example further demonstrated how a single interaction-based specification can give rise to both behavioral traces and structural interpretations without requiring external alignment or traceability mechanisms.

Across the analysis, SBC reframes several foundational concepts in systems theory. System identity is interpreted through interaction-based composition rather than component aggregation. Emergence is understood as a consequence of organized interaction rather than as an additional explanatory layer. Structural and behavioral representations are unified as projections of a single interaction-based system description.

From this perspective, Structure–Behavior Coalescence is not merely a modeling convenience, but a proposed reformulation of how systems are represented and understood in General Systems Theory [1–3]. It suggests that many of the difficulties in multi-view systems modeling may arise from representational separation rather than inherent properties of systems themselves [4–11].

While the present formulation is intentionally minimal, it establishes a coherent foundation for further development. Future work may extend the formal semantics of interaction-based composition, explore automated derivation of system views, and investigate applications in complex socio-technical and cyber-physical systems.

In summary, this paper advances a simple but foundational claim: a system is not what it is made of, nor what it does, but the organized interaction-based process through which structure and behavior coalesce.